\documentclass[12pt]{article}

\usepackage{graphicx}

\newcommand{\citep}{\cite}
\newcommand{\citet}{\cite}
\def\@cite#1{\mbox{$\m@th^{\hbox{\@ove@rcfont#1)}}$}}

\begin{document}


\begin{center}
{\large\bf
Anomalous 
fluctuations in sliding motion of cytoskeletal filament 
driven by molecular motors: 
Model simulations
}
\vskip 2ex

{\bf Yasuhiro$^1$ Imafuku, Namiko Mitarai$^2$,
Katsuhisa Tawada$^1$, and Hiizu Nakanishi$^2$}
\vskip 1ex

{\it 
$^1$ Department of Biology, Kyushu University, Fukuoka 812-8581, Japan \\
$^2$ Department of Physics, Kyushu University, Fukuoka 812-8581, Japan}

\end{center}

\begin{abstract}
It has been found in {\it in vitro} experiments that cytoskeletal
filaments driven by molecular motors show finite diffusion in sliding
motion even in the long filament limit [Y. Imafuku {\it et al.},
Biophys. J. 70 (1996) 878-886; N. Noda {\it et al.}, Biophys. 1 (2005)
45-53].  This anomalous fluctuation can be an evidence for 
cooperativity among the motors in action because fluctuation should be
averaged out for a long filament if the action of each motor is independent.
In order to understand the nature of the fluctuation in molecular
motors, we perform numerical simulations and analyse velocity
correlation in three existing models that are known to show some kind of
cooperativity and/or large diffusion coefficient, i.e.  Sekimoto-Tawada
model [K. Sekimoto and K. Tawada, Phys. Rev. Lett. 75 (1995) 180], Prost
model [J. Prost {\it et al.}, Phys. Rev. Lett.  72 (1994) 2652], and
Duke model [T. Duke, Proc. Natl. Acad. Sci. USA, 96 (1999) 2770].
It is shown that Prost model and Duke model do not give a finite
diffusion in the long filament limit in spite of collective action of
motors. 
On the other hand, Sekimoto-Tawada model has been shown to give the
diffusion coefficient that is independent of filament length, but it
comes from the long time correlation whose time scale is proportional
to filament length, and our simulations show that such a long
correlation time conflicts with the experimental time scales.
We conclude that none of the three models do not represent
experimental findings.  In order to explain the observed anomalous
diffusion, we have to seek for the mechanism that should allow both
the amplitude and the time scale of the velocity correlation to be
independent of the filament length.
\end{abstract}

\noindent {\bf keywords:} molecular motor, sliding motion, anomalous
fluctuation, cooperativity, model simulation

\newpage
\section{Introduction}

One of the outstanding problems in molecular processes in living systems
has been how they achieve reliable action under the influence of
overwhelming thermal and/or statistical fluctuations.  In the case of
muscle contraction, Huxley\citet{Huxley} has already noticed, in his
original work, 
that collective action of many motors produces smooth
sliding motion even though action of individual motor is highly
stochastic.

Recently, Imafuku and co-workers have done series of experiments on
filament motion driven by many molecular motors; Focusing on the {\it
fluctuation} rather than the average motion, they have revealed an
intriguing aspect of co-operativity in the collective
action\citep{Imafuku96a,Imafuku96b,Imafuku97,Noda}.

They performed the
{\it in vitro} motility assay on an unloaded filament of length $L$,
measured the displacement $X( t )$ over the time interval
between $t_0$ and $t_0+ t $, and evaluated the diffusion
coefficient $D$ in the sliding motion defined by
\begin{equation}
D= \lim_{t\to\infty}
\frac{\Bigl< \bigl( X( t )-\langle X( t )\rangle \bigr)^2\Bigr> }
{2t},
\label{Diffusion}
\end{equation}
where $\langle \cdots\rangle$ denotes the average over initial time
$t_0$ and samples.  If each motor exerting the force to the filament is
statistically independent, the fluctuation is averaged out as the
filament length $L$ becomes longer, namely, as the number of motors that
interact with the filament becomes larger.  It has been shown that $D$
decreases in proportion to $1/L$ for random action of independent
motors\citep{Sekimoto01}.  In the experiments, however, they have found
that $D$ is not proportional to $1/L$ but converges to a constant value
for large $L$.  This means that the motors are not interacting with the
filament independently but their actions are correlated with each other.

In order to understand these results, Sekimoto and Tawada have analyzed
the motion of a cytoskeletal filament driven by protein motors with random
orientation, and demonstrated that the diffusion coefficient $D$ of its
motion is independent of the filament length due to the randomness
quenched in the motor orientation\citep{Sekimoto95}.  Later, however,
Noda {\it et al} have found similar behavior of $D$ even for the case
where Sekimoto and Tawada model is not applicable, i.e. the case where
the myosins are not random but aligned\citep{Noda}.

No other model has been known to show the constant $D$ so far, and
origin of the observed fluctuations has not been understood yet.

In this paper, we study fluctuations of cytoskeletal filament motion in
detail for three existing models: Sekimoto and Tawada
model\citep{Sekimoto95}, Prost
model\citep{Prost,Julicher,Julicherreview,Badoual}, and Duke
model\citep{Duke}; These are known to produce some kind of
co-operativity\citep{Vermeulen02}.  As a tool to analyze dynamics, we
examine {\it the velocity-correlation function} of the filament sliding
motion obtained by numerical simulations.  Nature of dynamics shows up
in detailed feature of the velocity correlation function, and the
diffusion coefficient can be derived from its integration.

In Sec.2, some of the basic formulas are introduced in connection with
the diffusion coefficient and the velocity correlation.  Detailed
description and results for each model are presented with discussions
for Sekimoto-Tawada model in Sec.3, for Prost model in Sec.4, and for
Duke model in Sec.5.  Concluding remarks are given in Sec.6.


\section{Diffusion Coefficient and Velocity Correlation}

Imafuku {\it et al.}\citep{Imafuku96a,Imafuku96b,Imafuku97} measured the
variance of positional fluctuation of the filament defined as
\begin{equation}
F_r^2( t )\equiv 
   \Bigl< \bigl( X( t )-\langle X( t  )\rangle \bigr)^2\Bigr> , 
\label{variance}
\end{equation}
where $X( t )$ is the displacement over the time interval of length
$t$.  The average displacement $\langle X( t )\rangle $ is linear
in time, and the mean velocity $V$ of the filament is determined by
\begin{equation}
V=\lim_{t\to\infty}\frac{\langle X( t )\rangle}{t} .
\end{equation} 

They found that $F_r^2( t )$ increases linearly in time.  This is
an ordinary diffusion process, and is characterized by the diffusion
coefficient $D$ defined by
\begin{equation}
D \equiv \lim_{t\to\infty}\frac{F_r^2( t )}{2t} .
\label{dmdef}
\end{equation}
In the actual experiments, it has been found that
$F_r^2(t)$ behaves as
\begin{equation}
F_r^2(t) \approx 2D  t +\sigma,
\label{dmdef2}
\end{equation}
with a constant $\sigma$, which mainly comes from the finite spatial
resolution in experiments\citep{Imafuku96a}.

The diffusion around the average motion comes from the velocity
fluctuations in the sliding motion, and $F_r^2( t )$ can be
expressed as
\begin{equation}
F_r^2( t )=2  t  \int_0^{ t }
\left(1-\frac{s}{ t }\right) C_v(s) {\rm d}s
\label{eq:fr2cv}
\end{equation}
in terms of the velocity correlation
\begin{equation}
C_v( t )=\Bigl< \bigl( v(t_0)-V\bigl)\bigr( v(t_0 +t)-V\bigr)\Bigr> ,
\label{C_v}
\end{equation}
where $v(t)$ is the velocity at time $t$ and $V$ is the
average velocity.  The derivation of eq.(\ref{eq:fr2cv}) 
is given in Appendix.

Note that the relation (\ref{eq:fr2cv}) is modified when the data is
only available at discrete times by the step $\tau$ as in the
experiments or our Monte Carlo (MC) simulations: In this case, defining of
velocity at time $t_j=\tau j$ as $v(t_j)=[X(t_j+\tau)-X(t_j)]/\tau$, the
diffusion coefficient $D$ is given by
\begin{equation}
D=\left[\frac{1}{2}C_v(0)+\sum_{j=1}^{\infty}C_v(t_j)\right]\tau,
\label{eq:discreteDm}
\end{equation}
in the long-time limit(see Appendix).

The velocity correlation function $C_v(t)$ goes to zero when $t$ becomes
large enough compared to any relevant correlation time, therefore, the
second term contribution in eq.(\ref{eq:fr2cv}) becomes negligible in
the large $ t $ limit, consequently, $F_r^2(t)$ increases linearly in
time and $D$ in  eq.(\ref{dmdef}) can be expressed by
\begin{equation}
D=\int_0^{\infty} C_v(s)\mbox{d}s,
\label{eq:dminteg}
\end{equation}
namely, the diffusion coefficient is given by the integral of the
velocity correlation function.  

Actual measurements are always based on finite time observations, thus we
define the finite time diffusion coefficient $D_{\rm ft}(t)$ as the slope
of $F_r^2( t )$ at $t$, then we can show
\begin{equation}
D_{\rm ft}( t ) \equiv
\frac{1}{2} \frac{\rm d}{{\rm d} t }F_r^2( t )
=
\int^{ t }_0 C_v(s){\rm d}s,
\end{equation}
namely, only the correlation shorter than $t$ contributes to
$D_{\rm ft}(t)$.

Experimentally determined diffusion coefficient based on the measurement
of variance for the time interval $t$ corresponds to $D_{\rm ft}(t)$.  This
should give a good approximation for the diffusion coefficient
(\ref{eq:dminteg}), if $t$ is large enough compared to the correlation
times.  We should be careful about the effect of the measurement time $
t $ when we interpret the results, because we do not know the length of
correlation time for the system in advance.
Imafuku and coworkers estimated the diffusion coefficient $D$ from
experimental data by the slope of $F_r^2(t)$ at around $t \approx
2\,{\rm s}$ for microtubules driven by kinesin\citep{Imafuku96b} $t
\approx 0.5\,{\rm s}$ for microtubules driven by
dynein\citep{Imafuku97}, and $t \approx 0.4\,{\rm s}$ for actin
filaments driven by myosin\citep{Noda}.

In the following, we analyze the diffusion constant $D$ along with the
velocity correlation.

\section{Sekimoto and Tawada Model}


Sekimoto and Tawada have proposed a simple model to explain dynamical
fluctuations in the motion of a cytoskeletal filament driven by protein
motors fixed on a substrate surface\citep{Sekimoto95}; The motor
proteins are assumed to be aligned at regular intervals $q$ but in
random orientation, and a filament of the length $L$ slides over the
motors in one direction in the presence of ATP.  The sliding motion
is generated by the motors, which are assumed to attach to the filament
with the rate constant $k_{\rm b}$, make a conformational change or
``power stroke'' to generate force, and detach from the filament with
the rate constant $k_{\rm ub}$.  The motors interact with
the filament independently.  If the linear force law between the
filament and the motors is employed, the sliding distance generated by a
power stroke of the width $a_i$ by the $i$'th motor would be $a_i/N_b$,
where $N_b$ is the number of the motors that attach to the filament at
the time of the stroke.  The factor $N_b^{-1}$ comes from the fact that
the motors bound to the filament at the time resist the sliding motion.
The value of stroke width $a_i$ for the $i$'th motor is always the same,
but different at random from that of other motors because of its random
orientation.

Sekimoto and Tawada have demonstrated that the model shows $L$
independent diffusion constant.

\subsection{Simulation method}

We simulate Sekimoto-Tawada model by the following 
MC procedure with the time $\tau$ for one MC step:
(a) Pick a motor, say $i$, at random out of the motors under the
filament.
(b) If the $i$'th motor is not attached to the filament, attach it with
the probability $k_{\rm b} \tau$, and advance the filament by
$a_{i}/N_b$.  If the $i$'th motor is already attached to the filament,
detach it with the probability $k_{\rm ub} \tau$ without any
motion of the filament.

One MC step of $\tau$ corresponds to repeating this procedure $N$ times,
where $N$ is the number of motors that are capable of attaching to the
filament: $N=L/q$.  The results do not depend on $\tau$ as long as $\tau
k_{\rm b}\ll 1$ and $\tau k_{\rm ub}\ll 1$.

We adopt the rate constant of the motor attachment $k_{\rm b}=6.3\,\rm
s^{-1}$, that of the detachment $k_{\rm ub}=$14.7 $\rm s^{-1}$, and the
distance between motors $q=42.9\,{\rm nm}$.  The width of the power stroke
$a_i$ is chosen out of random numbers with the uniform distribution
ranging between $0$ and $34$ nm with the average $17\,{\rm nm}$.  The time
step $\tau$ is taken to be $0.01\,{\rm s}$.

\subsection{Simulation results}
\begin{figure}
\centerline{\includegraphics[width=0.5\textwidth]{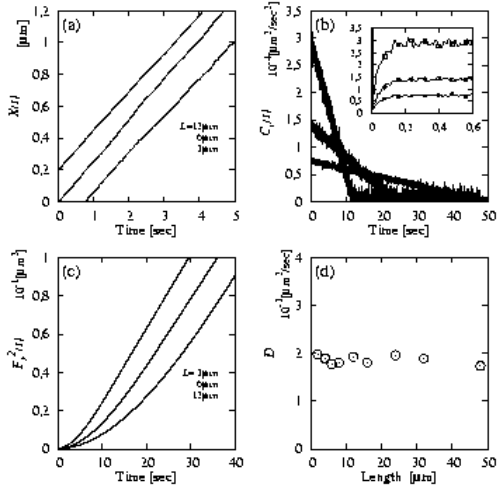}}
\caption{
Sliding motion of Sekimoto-Tawada model.
(a) Typical trajectories of filaments with the length
$L=$3(bottom), 6, and 12(top) $\mu$m.
The average velocity $V=0.25 \mu{\rm m/s}$ is 
independent of the filament length $L$.
(b) Velocity correlation $C_v(t)$
for filaments with $L=$3, 6, and 12 $\mu$m. 
The inset shows short time behaviors ($0< t\le 0.6$ s).
%
%
(c) Variance as a function of time interval for
 $L=$3(top), 6, and 12(bottom) $\mu$m. 
(d) Length dependence of the 
diffusion coefficient $D$ measured from the slope for
$L/V$ $\le  t \le 2L/V$.
%
}
\label{Fig2}
\end{figure}

Figure \ref{Fig2}(a) shows typical trajectories of filaments with
$L=$3, 6, and 12 $\mu$m.  The average velocity $V$ is $0.25 \mu$m/s,
and is independent of $L$.

Figure \ref{Fig2}(b) shows the velocity correlation $C_v(t)$ for rather
long time ($0<t\le 50\,{\rm s}$), and the inset shows the short time
behaviors for $0<t\le 0.6$ s.  We find following characteristic behavior
in $C_v(t)$:
(i) At $t=0$,
the velocity correlation $C_v(0)$ is the variance of the
velocity fluctuation, which takes a certain positive value.
%
%
(ii) For $t>0$, $C_v(t)$ drops immediately to almost zero, 
but recovers quickly to reach the maximum.
(iii) After the short time recovery, $C_v(t)$ decays linearly over
a long period of time.
(iv) The value of $C_v(t)$ scales as to $1/L$ for both $t=0$ and the
maximum value described at (ii).

Note that the actual values of instantaneous velocity variance
$C_v(0)$ do not have physical meaning because the model assumes
instantaneous displacement by a stroke of molecular motor.

The immediate drop of $C_v(t)$ comes from the fact that the
consecutive strokes to the filament is random.  The short time recovery
is due to the correlation given by the strokes from the same motors.
The time $T$ when the maximum correlation is achieved corresponds to the
time interval that each motor gives successive strokes to the filament,
namely $T\approx 1/k_{\rm b}+1/k_{\rm ub}(\approx 0.23\, {\rm s})$.
This time does not depend on $L$.  The gradual linear decay after that
comes from the loss of correlation due to the fact that new motors come
into play as the filament moves.  The correlation is lost completely
when the filament proceeds over the distance $L$ and resides on a new
set of motors, therefore, the time scale for this correlation is $L/V$.

The variance of displacement $F_r^2(t)$ is shown in Fig.\ref{Fig2}(c)
by a solid line for $L=$3, 6, and 12$\,\mu$m.  
We see that the slope increases gradually
in course of time till $t \approx L/V$.  This comes from the slow
decay of the velocity correlation $C_v(t)$ in Fig.\ref{Fig2}(b) through
Eq.(\ref{eq:fr2cv}), namely, the slope at time $t$ is given by
the integration of $C_v(s)$ for $0\le s \le t$.  The slope of the
variance is proportional to $1/L$ for a fixed short time $t \ll L/V$,
because $C_v(t)\propto 1/L$ for a fixed time $t<L/V$.  The slope for
long time $t > L/V$, however, becomes independent of $L$, because the
range that $C_v(t)$ is non-zero is proportional to $L$.

Fig.\ref{Fig2}(d) shows the diffusion coefficient $D$ obtained from the
slope for long time (fitted in the range $L/V\le t \le 2L/V$).  One can
see that $D$ tends to a non-zero constant for larger $L$.  As denoted
above, however, $D$ should be proportional to $1/L$ if one measures at
fixed $t \ll L/V$ in this model.

\subsection{Discussions}

The key of this model is that each motor gives strokes of its inherent
strength, simulating the experimental situation where motors are fixed
to a substrate surface in random orientation and are not aligned to the
direction of the filament; Each motor always gives the same stroke,
although it varies from motor to motor.  This makes the correlation time
in velocity fluctuation proportional to $L$, and results in the $L$
independent diffusion constant.  If a stroke $a_i$ by a motor changes
randomly every time it gives a stroke, the correlation time is
independent of $L$, thus the diffusion coefficient $D$ would be
proportional to $1/L$.

This assumption of quenched randomness in stroke in the Sekimoto-Tawada
model is introduced in order to simulate the experimental situations by
Imafuku {\it et al.}\citet{Imafuku96b,Imafuku97}, where microtubules are
driven by motors scattered randomly on a substrate.
Their results that the diffusion coefficient in sliding motion of
microtubules is independent of $L$ are reproduced by the model.

The present analysis shows, however, that this does not necessarily
justify the model because the experimental time scales are not in the
range where the model gives $L$ independent diffusion coefficient.  In
these experiments, $L/V$ was $0.5\,{\rm s}$ to $2\,{\rm s}$ in
Ref.\citet{Imafuku96b} and $0.6\,{\rm s}$ to $2.5\,{\rm s}$ in
Ref.\citet{Imafuku97}, while the time $t$ of the measurement was about
$2\,{\rm s}$ in Ref.\citet{Imafuku96b} and $0.5\,{\rm s}$ in
Ref.\citet{Imafuku97}, respectively.  Namely, the latter experiment by
Imafuku {\it et al.}\citet{Imafuku97} were in the range $t< L/V$, i.e.,
the range where Sekimoto-Tawada model shows the diffusion coefficient
that decreases with $L$. Therefore, the results of the model does not
correspond to the experimental observation by Imafuku {\it et
al.}\citet{Imafuku97}.

Relevance of quenched randomness to the anomalous diffusion is also
questionable; Kinesin has been demonstrated to swivel almost freely to
adjust itself in any direction to give effective strokes\cite{Hunt1993}.
Furthermore,
the anomalous filament length independent diffusion has been observed
even in the experiment where myosins are aligned in
orientation\citet{Noda}.

\section{Prost Model}

\begin{figure}
\centerline{\includegraphics[width=0.4\textwidth]{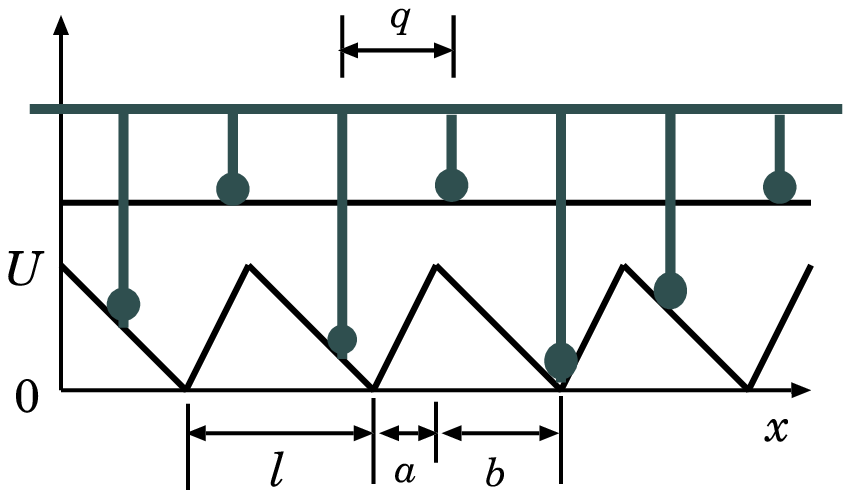}}
\caption{Prost Model}
\end{figure}

Prost and coworkers introduced a two-state model of Brownian motor,
which exhibits cooperative behavior\citep{Prost,Julicherreview}.  In
their model, motors are attached to a backbone with a fixed spacing $q$,
and each motor can take two states: the bound state and the unbound
state.
In the bound state, the motor is strongly bound to a filament to exert
the sliding force on it, and the interaction potential between the
filament and the motor is periodic with period $l$ along the cytoskeletal
filament.  A simple saw teeth potential is assumed; Each tooth of the
potential height $U$ consists of a part with a positive slope of length
$a$ and a part with a negative slope of length $b\equiv l-a$.
The unbound state represents the weakly
bound state, where the interaction potential is flat, and the
motor does not exert force on the filament.
The transition from the bound to the unbound state takes place with the
rate $\omega_{\rm ub}$ only when a motor is around the potential minima
within a detaching region of size $\delta x$.  The transition from the
unbound to the bound state, on the other hand, occurs everywhere with a
constant rate $\omega_{\rm b}$.  The inertia of the filament is assumed
to be negligible, thus the instantaneous velocity of the filament $v$ is
determined by the balance between the total potential force from the
motors $F_{\rm mot}$ and viscous force $-N\lambda v$, where $N$ is the
number of motors and $\lambda$ is the viscous resistance coefficient per
motor.

This model has been demonstrated to show a collective motion of smooth
sliding when a number of motors are attached to a
filament\citep{Badoual}, in contrast with a Brownian motion under a
periodic potential for a single motor system.  If the periodic potential
is asymmetric, the filament shows unidirectional motion, while
bidirectional motion is observed when the potential is symmetric.

The cooperativity of the motors is evident especially in the symmetric
case.  In this case, there is no reason for the filament to proceed in one
direction, but once the filament starts moving to one of the directions
by chance, it tends to keep on moving in the same direction.  This
can be understood from the fact that the transition from the bound to the
unbound state occurs only when the motors are around the potential
minima; The motors are kicked out to the unbound state after they go
down the potential slope exerting the force to the filament, therefore,
there are more motors going down the potential than those going up in
the bound state.  This state of directed motion may be regarded as a
state with a broken symmetry due to the collective 
effect\citep{Badoual}.

The direction of motion flips occasionally when the filament length is
finite, and the interval between the flips gets exponentially long with
the filament length.  A theory analogous to the one for the phase
transition in a magnetic system has been developed to describe this
state.

\subsection{Simulation method}


One Monte Carlo step with the time $\tau$ consists of the two
procedures: (a) the transition trials between the bound and the
unbound states of motors, and (b) the filament advance.
In the procedure (a), each of the $N$ motors goes through the
following transition trial depending upon its state: if the motor is
in the bound state and it is located in one of the detaching regions
of size $\delta x$ around the potential minima, detach it to the
unbound state with the probability $\omega_{\rm ub}\tau$.  If the
motor is in the unbound state, attach it to the bound state with the
probability $\omega_{\rm b}\tau$. Otherwise, do nothing.
After all of the motors go through the above trials, 
we perform the procedure (b), where
 the total potential force $F_{\rm mot}$ acting on the motors in
the bound state is calculated, and then the filament is displaced by
$\tau v$ with the filament velocity $v=F_{\rm mot}/(N\lambda)$.

%
%
%
%
%
%


The period of potential is taken to be $l=8\,{\rm nm}$, and the motor
spacing $q=42.9\,{\rm nm}$ with small random distribution of the width
$\pm 0.01\,{\rm nm}$; The small distribution is introduced in order to
avoid factitious interference between the motor spacing and the
potential periodicity.  The potential is chosen to be $a/l=0.2$ and
$U=20 k_B T$ with $k_BT=4.14\, {\rm pN\,nm}$\citep{Julicher}.  The
transition rates are $\omega_{\rm ub}^{-1}=2$ ms and $\omega_{\rm
b}^{-1}=25$ ms, and the size of the detaching region is $\delta x=1.6$
nm.  We also adopt the viscous resistance coefficient
$\lambda=1.29\times 10^{-5}\,{\rm kg/s}$.

\subsection{Simulation results}

\begin{figure}
\centerline{\includegraphics[width=0.5\textwidth]{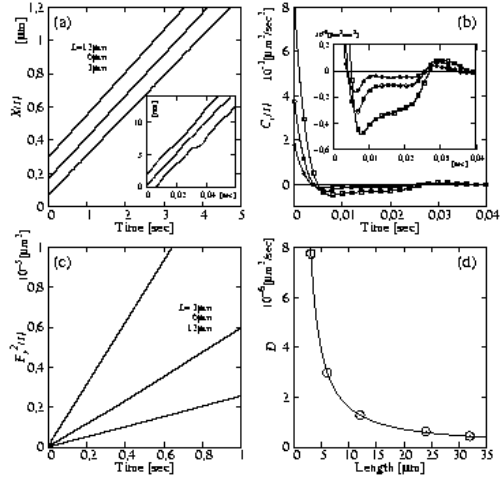}}
\caption{
Sliding motion of Prost model.  
(a) Typical trajectories of filaments with the length 
$L=$3(bottom), 6, and 12(top) $\mu$m.  The average velocity is almost
independent of the filament length $L$, but slightly smaller for shorter
filaments: $V=$0.239, 0.248, and 0.253 $\mu{\rm m/s}$ for $L=$3, 6,
 and 12 $\mu{\rm m}$, respectively.
(b) Velocity correlation $C_v(t)$ for filaments with $L=$3(open square),
6(open circle), and 12(filled circle) $\mu$m.  The inset shows the
behaviors for $t>0$ s in larger scale.
%
%
(c) Variance as a function of time interval for
$L=$3(top), 6, and 12(bottom) $\mu$m. 
(d) Length dependence of the 
diffusion coefficient $D$. The solid line shows a fitting curve 
$a/L\cdot(1+b/L)$
via constants $a=12.72\times10^{-6}\,\mu{\rm m^3/s}$ and $b=2.48\,\mu{\rm m}$.
} \label{fig2}
\end{figure}

Examples of the simulated displacements of motors versus time under no
load condition are shown for $L=$3, 6, and 12 $\mu{\rm m}$ in
Fig.\ref{fig2}(a).  The average velocity is almost independent of the
filament length $L$, but slightly smaller for shorter filaments:
$V=$0.239, 0.248, and 0.252 $\mu{\rm m/s}$ for $L=$3, 6, and 12 $\mu{\rm
m}$, respectively.  The detailed structure of trajectories are shown in
the inset, where one can see more fluctuations in a shorter filament.
Figure \ref{fig2}(b) shows the velocity correlation
$C_v(t)$, for which we note the following characteristics:
(i) $C_v(t)$ is largest at $t=0$, and decays rapidly to
reach the negative minimum around $t\sim 6$ ms.
(ii) Then, it increases gradually to reach the positive maximum at the
time $t\sim 30$ ms, and decays to zero as $t$ increases.
(iii) The value of $C_v(t)$ is proportional to the inverse of the length
of filament, or the number of motors.
In Fig. \ref{fig2}(c), the variance $F_r^2(t)$'s are shown as a function
of time.  The diffusion coefficient is evaluated from the time
dependence of the variance by fitting the data for $0\, {\rm s} <t<0.8\,{\rm
s}$ to Eq.(\ref{dmdef2}).
The obtained values of the diffusion coefficient $D$ versus the length
of the filament $L$ are shown in Fig. \ref{fig2}(d) with a fitting curve
$a/L\cdot (1+b/L)$.

\subsection{Discussions}

The velocity fluctuation in this model comes from random transition of
motors between the two states.  Consequently, the amplitude of velocity
correlation is proportional to $1/L$ for large $L$, reflecting the fact
that the transitions of each motor is independent.  The time scale of
the initial decay in $C_v( t )$ is set by the transition rate
$\omega$'s, and the correlation time of the positive peak in $C_v( t )$
at $t\sim 30$ ms is the time that a motor passes the period of the
potential ($l/v$ = 8 nm / 0.25 nm (ms)$^{-1} \sim 30$ ms), thus, both of
the time scales are independent of $L$ while $C_v\propto 1/L$,
therefore, the resulting diffusion constant for this model is
proportional to $1/L$ as is seen in Fig.\ref{fig2}(d).

\section{Duke Model}

Duke\citet{Duke} proposed a model for the myosin mechanochemical
cycle, based on the ``swinging lever arm'' hypothesis, and demonstrated
that the model shows collective behavior of myosins through coupling
of ATP hydrolysis with conformational change of myosin head.
The conformation change is amplified by a lever arm to produce a power
stroke.  The power stroke stretches the myosin neck and the actin
filament slides as the neck relaxes.

\begin{figure}
\centerline{\includegraphics[width=0.35\textwidth]{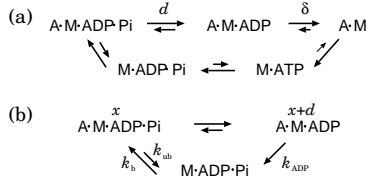}}
\caption{ATP hydrolysis cycles.}
\label{cycle}
\end{figure}

The ATP hydrolysis cycle is simplified as
in Fig.\ref{cycle} (a);
A main stroke of the step width $d$ takes place at the transition from
${\rm  A\cdot M\cdot ADP\cdot Pi}$ to ${\rm  A\cdot M\cdot ADP}$, then
a small stroke of the width $\delta$ follows at the subsequent
transition to ${\rm A\cdot M}$.

Eliminating a couple of transient states, i.e. ${\rm A\cdot M}$ and ${\rm
M\cdot ATP}$, this cycle is further simplified to the three-state cycle
shown in Fig.\ref{cycle} (b),
where the small step is included in the transition from ${\rm A\cdot
M\cdot ADP}$ to ${\rm M\cdot ADP\cdot Pi}$.

To each myosin in the attached state ${\rm A\cdot M\cdot ADP\cdot Pi}$,
Duke assigned the displacement $x$ of the neck, which becomes $x+d$ for
${\rm A\cdot M\cdot ADP}$ after a stroke.  The elastic energy of the
myosin neck at the state ${\rm A\cdot M\cdot ADP\cdot Pi}$ and that at
the state ${\rm A\cdot M\cdot ADP}$ are $Kx^2/2$ and $K(x+d)^2/2$,
respectively, with the spring constant of the neck $K$.

The binding rate $k_b$ from ${\rm M\cdot ADP\cdot Pi}$ to ${\rm A\cdot
M\cdot ADP\cdot Pi}$ with the neck displacement $x$ is affected by this
elastic energy;
The binding rate to the state with the neck displacement in the range
$[x,x+{\rm d}x]$ is given by
\begin{equation}
k_b\, {\rm d}x = k_b^0\,
\sqrt\frac{K}{2\pi k_B T}\, 
\exp\left[-\,\frac{Kx^2}{2 k_B T}\right] {\rm d}x,
\label{k-bind}
\end{equation}
while the unbinding rate $k_{\rm ub}$ for the opposite transition is
assumed to be constant and independent of $x$.

The transition between the pre and post stroke states, i.e.  $\rm A\cdot
M\cdot ADP\cdot Pi$ and $\rm A\cdot M\cdot ADP$, is fast and
the population ratio is assumed to
be equilibrated as the ratio of the probability distribution
\begin{equation}
\frac{P\left( {\rm A\cdot M\cdot ADP}\right)}
{P\left( {\rm A\cdot M\cdot ADP\cdot Pi}\right)}
=
\exp\left[
-\,\frac{\Delta G_{\rm str}+\Delta E(x)}{k_B T}
\right],
\label{equil-ratio}
\end{equation}
where $\Delta G_{\rm str}$ is a change in chemical free energy and
\[
 \Delta E(x) \equiv \frac{1}{2} K(x+d)^2 - \frac{1}{2} Kx^2
\]
is a change in elastic energy.

Within the transition from $\rm A\cdot M\cdot ADP$ to $\rm  M\cdot ADP\cdot
Pi$, series of transitions, including $\rm ADP$ release with the small step
$\delta$, $\rm ATP$ bind followed by actin detachment, are involved, and
its overall rate is represented as
\begin{equation}
k_{\rm ADP} = k_{\rm ADP}^0 
      \exp\left[ -\,\frac{\Delta E_\delta(x)}{k_B T} \right]
\end{equation}
with
\begin{equation}
\Delta E_\delta(x) \equiv 
            \frac{1}{2} K(x+d+\delta)^2 - \frac{1}{2} K(x+d)^2  ,
\end{equation}
but the transition in the opposite direction is neglected.

The time scale for the force balance is assumed to be much shorter than
the time scale for the transition between the states, thus the position
of a filament is always adjusted to achieve the force balance among the
molecular motors as soon as one of the motors changes its state.


Duke investigated how an ensemble of motors generates sliding motion
of an actin filament against load.  The model displays a transition from
smooth sliding to synchronized stepwise motion as the load becomes high.

\subsection{Simulation method}

Suppose an actin filament of the length $L$ is sliding straight on a
substrate under load.  The myosin molecules are set in array on the
substrate with spacing $q$, thus the number of the myosins that can
possibly attach is $N=L/q$.

In each MC step of the time $\tau$, the following two procedures are
iterated $N$ times: 
(a) transition between attached and detached states
and 
(b) equilibration of attached molecules.
Detailed procedures are implemented as follows;
(a) Transition goes by two steps.
(a-1) Pick a motor at random. 
(a-2) If the motor is not attached to the filament, then attach it with
the probability $k_{\rm b}^0 \tau$ with its neck displacement $x$ chosen
at random according to the distribution (\ref{k-bind}).
If the motor is attached, detach it with the probability $k_{\rm
ub}\tau$ or $k_{\rm ADP}\tau$, depending upon it is in 
$\rm A\cdot M\cdot ADP\cdot Pi$ or $\rm A\cdot M\cdot ADP$,
respectively.
(b) Equilibration is achieved by repeating the following steps $M$
times.
(b-1) Move the filament to the position where the force from the motors
balances with the load.
(b-2) Distribute all the attached myosins according to the population ratio
(\ref{equil-ratio}).

The procedures (a) and (b) are repeated
$N$ times in one MC step of time $\tau$; Within (b), the procedure (b-1) and
(b-2) are repeated $M$ times to ensure that the system
position is relaxed to equilibrium\cite{com1}%
%
.

The parameters used in the simulations are followings: the spacing
between myosin molecules is $q=42.9$ nm, the rate constant of the motor
attachment $k_{\rm b}^0 = 40\,{\rm s^{-1}}$, that of the detachment
$k_{\rm ub} = 2\,{\rm s^{-1}}$ and $k_{\rm ADP}^0 = 80\,{\rm s^{-1}}$,
the power stroke size $d = 11$ nm, the stiffness of myosin neck
$K=1\,{\rm pN/nm}$, the thermal energy $k_BT=4.14$ pNnm, the free energy
gain $\Delta G_{\rm str}=-16.4 k_BT$, and the subsequent conformational
change $\delta$ =0.5 nm. The time step $\tau$ is taken to be 0.001 s and
the equilibration iteration $M$ to be 4.

\subsection{Simulation results}
\begin{figure}
\begin{center}
\includegraphics[width=0.5\textwidth]{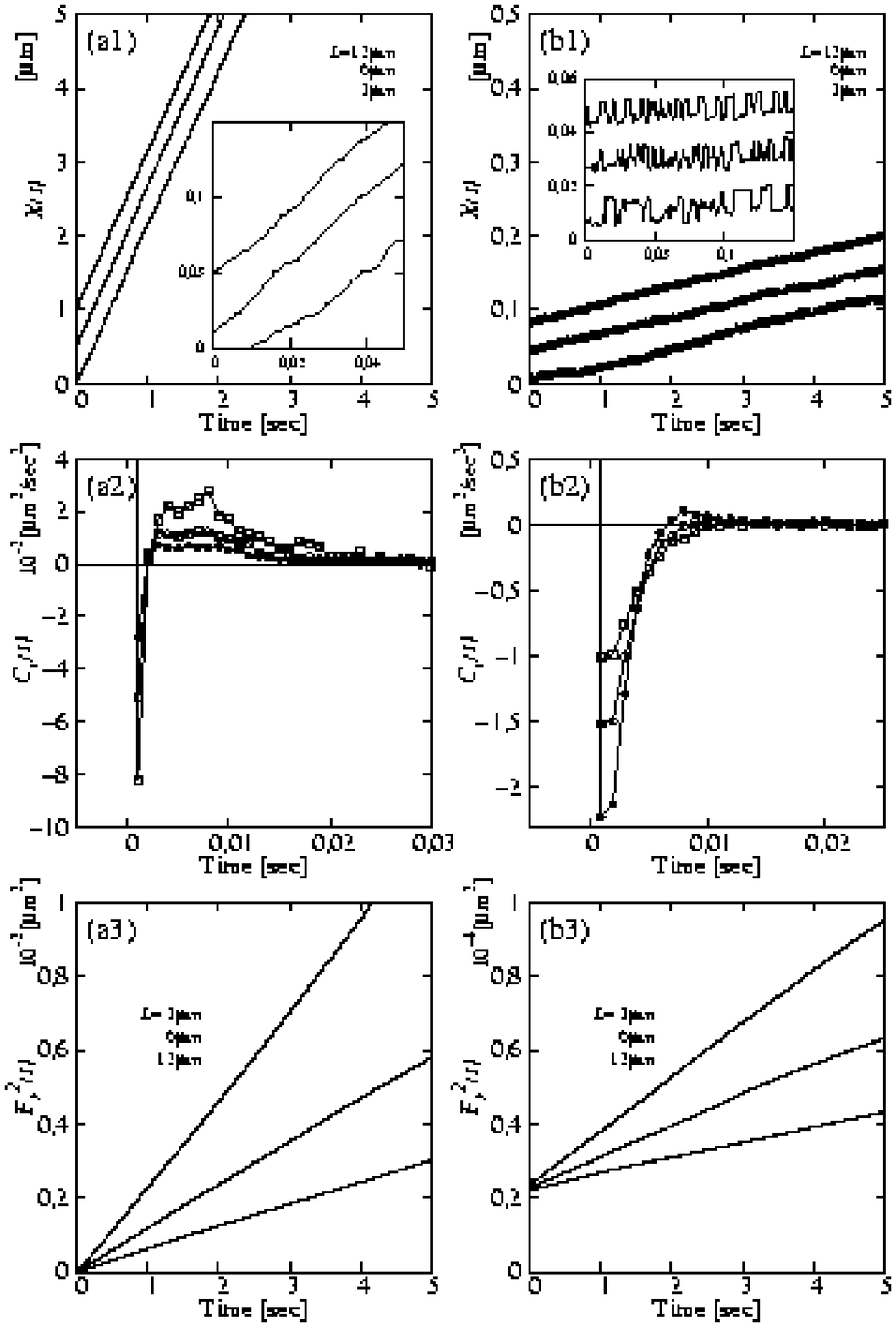}\\
\includegraphics[width=0.5\textwidth]{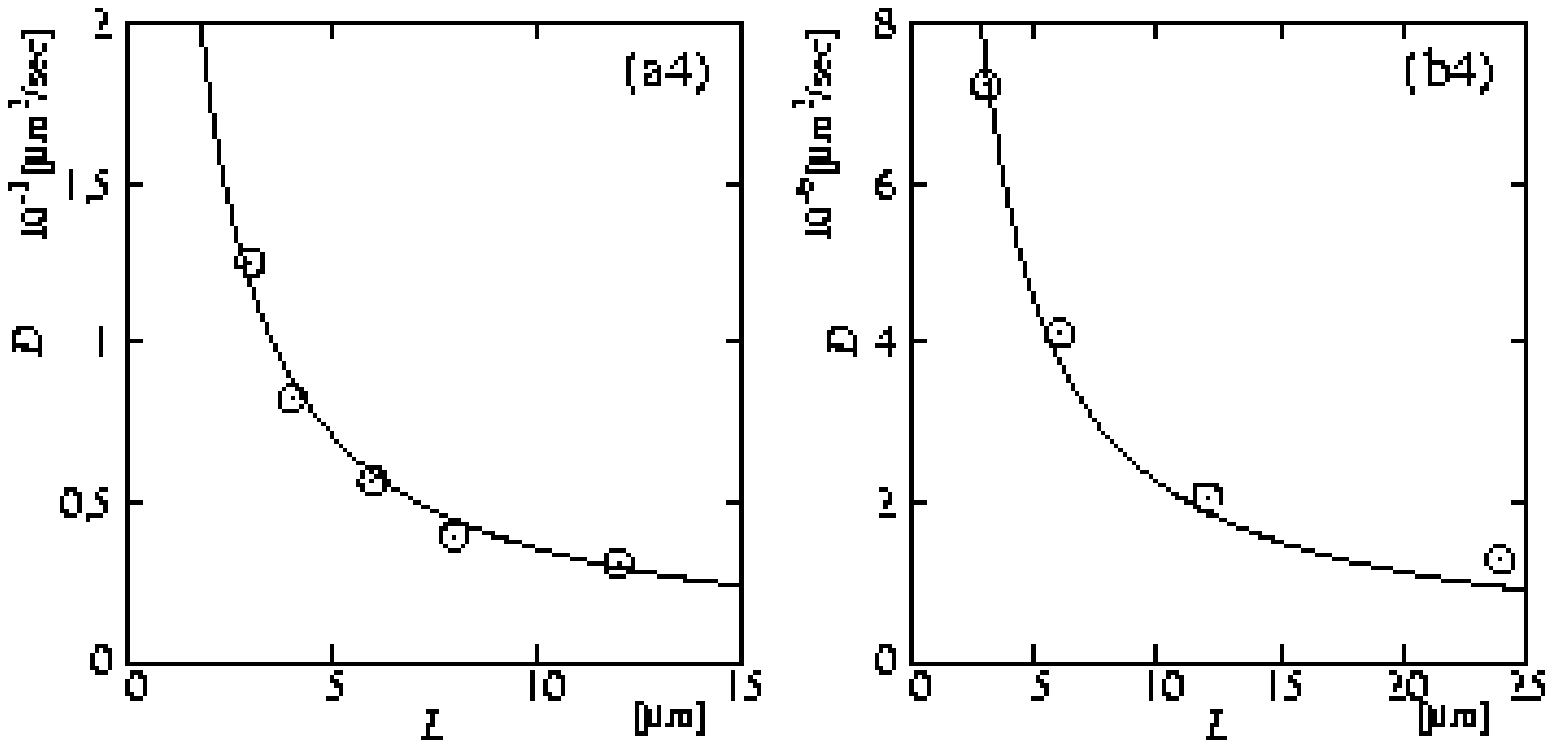}
\end{center}
\caption{
Sliding motion of Duke model
(a1-4) for cases without load and (b1-4) with load of 100 pN/$\mu$m.
(a1) and (b1) show 
typical trajectories of filaments with the length
$L=$3(bottom), 6, and 12(top) $\mu$m.
The average velocity $V$ is almost independent of $L$:
$V=$2.06, 2.11, and 2.14 $\mu{\rm m/s}$ without load, and
$V=$1.98, 2.25, and 2.36 $\times 10^{-2} \mu{\rm m/s}$ with load.
The insets show detailed structure of trajectories.
(a2) and (b2) show the velocity correlation $C_v(t)$ for filaments with
$L=$3(open square), 6(open circle), and 12(filled circle) $\mu$m.
(a3) and (b3) show the variances as a function of time interval for
$L=$3(top), 6, and 12(bottom) $\mu$m. 
(a4) and (b4) show the length dependence of the 
diffusion coefficient $D$ determined by the slope of variances.
The solid lines show the $1/L$ fit to the data points.
}\label{fig3}
\end{figure}

We show two sets of data in Fig.\ref{fig3}: the data for the case
without load (Fig.\ref{fig3}(a1-4)) and with the load $F$ close to
stalling, i.e.  $F=$100 pN/$\mu$m(Fig.\ref{fig3}(b1-4)).  We examine the
behavior near the stalling load because cooperative operation of motors
has been observed under load\citep{Duke}, which could
result in fluctuations peculiar to the model.  For each case, we present
trajectories, velocity correlations, time developments of displacement
variance, and the filament length dependences of diffusion constant.

Figures \ref{fig3}(a1) and (b1) show trajectories for filaments with
$L=$3, 6, and 12$\mu{\rm m}$.  The average velocity $V$ is almost
independent of $L$: $V=$2.06, 2.11, and 2.14 $\mu{\rm m/s}$ without
load, and $V=$1.98, 2.25, and 2.36 $\times 10^{-2} \mu{\rm m/s}$ with
load.  The insets show detailed structure of trajectories.  One can see
larger fluctuations in shorter filaments.  For the cases under the load,
there appear back and forth movements(the inset of Fig.\ref{fig3}(b1));
Displacement of the movement is of the order of a stroke size $d$, and
its time scale is determined by the reaction rates.  This is a result of
cooperative action of motors; With more motors attached, they produce
force collectively to move forward, but detachment of some motors causes
backlash, which leads to further detachment.

The velocity correlations are shown in Figs.\ref{fig3}(a2) and (b2).
For both of the cases, immediately after large positive instantaneous
correlation, the correlation is negative for short time and becomes
positive around $t=0.01$ s.  The initial negative correlation is due to
the backlash caused by detachment after a stroke.  The time scale for
correlation is determined by the reaction rates and does not depend on
the filament length.  Note that the actual value of instantaneous
correlation is not meaningful in this model as in Sekimoto-Tawada model,
because filament position is shifted instantaneously during the process
of equilibration.
%

Difference between the two cases is in the amplitude of the correlation.
First, the amplitude is much larger for the cases with load than those
without load.  The large correlation in the case under load comes from
the back and forth movement.  Secondly, regarding the $L$ dependence,
the amplitude is larger for shorter filament for the loadless case as in
the previous model, while it is smaller for shorter filament for the
case with load.  The larger correlation for shorter filament in the
loadless case corresponds to the fact that the fluctuations in
trajectories are larger for shorter filaments as has been seen in
Fig.\ref{fig3}(a1).  On the other hand, the larger correlation for
longer filament in the case with load corresponds to the fact that the
short time back-and-forth movement is more regular in longer
filament(the inset of Fig.\ref{fig3}(b1)).  The fluctuation in the time
scale longer than 1 second seems to be larger in shorter filament for
the case with load, but the correlation in such long time scale cannot
be seen in Fig.\ref{fig3}(b2) because of statistical errors.

The time dependences of displacement variance are shown in
Figs.\ref{fig3}(a3) and (b3), and the diffusion constant estimated from
these are presented in Figs.\ref{fig3}(a4) and (b4) with the $1/L$
curves fitted to the data. For both cases, the diffusion constant is
proportional to $1/L$.  This is natural for the case without load,
because the time scales of $C_v(t)$ are independent of $L$ while the
value of $C_v(t)$ is proportional to $1/L$.  In the case with load, the
large value of the correlation does not give a large value of $D$; This
means that the back and forth movement in short time scale does not give
net motion, and the diffusion comes from the fluctuation in longer time
scale, where the fluctuation is larger for shorter filament even in the
case with load.

\subsection{Discussions}

The diffusion constant decreases as $1/L$ with the filament length $L$.
In the case without load, the situation is simple; the time scale of
velocity correlation is independent of $L$ and its value is proportional
to $1/L$, which gives $D\propto 1/L$.

In the case with load, the motors on a filament operate collectively,
which results in back and forth movement in short time scale.  This back
and forth movement becomes more regular and gives larger value of
correlation for longer filament, but does not results in larger
diffusion.  The diffusion in the sliding motion comes from longer time
fluctuations, which is larger for shorter filament.


\section{Concluding Remarks}

Our results of numerical simulations are summarized as follows:
(i) Although the amplitude of velocity correlation function decreases as
$1/L$, Sekimoto-Tawada model shows the $L$ independent diffusion
coefficient $D$ because the time scale of velocity correlation is given
by $L/V$, thus proportional to $L$.
(ii) Prost model shows $D\propto 1/L$ because the amplitude of velocity
correlation is proportional to $1/L$ and its time scale is independent
of $L$.
(iii) Duke model also shows $D\propto 1/L$ as in Prost model.


Although the anomalous diffusion in the experiments is observed
without load, we examined Duke model with load also, because Duke
model has been denmonstrated to show a synchronous operation under
load and we expected that filament motion caused by collective
operation of motors was a possible explanation for large diffusion.
However, the collective movement by synchronous operation under load
in Duke model did not turn out to produce large diffusion for a long
filament; It gives rather regular back and forth movement in short
time, but did not give a net motion for longer time.


In connection with the experimental observation that $D$ is finite in
the large $L$ limit, the $L$ independent $D$ in Sekimoto-Tawada model is
not likely to be relevant; In the experimentally determination of
diffusion coefficient, only the time dependences of the variance shorter
than a few seconds are used, which means that the velocity correlation shorter
than a few seconds is relevant, while the correlation time in Sekimoto-Tawada
model is of order of $L/V$, which can be much longer than the
experimental time scale for a long filament.

In conclusion, by examining the velocity correlation and the diffusion
coefficient, we have confirmed the existing models cannot account for
the anomalous fluctuation that the diffusion coefficient remains finite
even in the large $L$ region.  Our analysis shows that the experimental
observations of $L$-independent diffusion coefficient require a
mechanism that makes both the amplitude and the time scale of velocity
correlation function independent of $L$.


\appendix
\section{Relation between Variance and 
Velocity correlation Function}

We briefly summarize the derivation 
for some formulas which express the variance $F_r^2(t)$ in terms of
the velocity correlation function $C_v(t)$
following Ref.\citet{Hansen} for both the continuous and discrete time data.

\subsection{Continuous time expression}

The position of the filament $X(t)$ at time $t$ is given by
\begin{equation}
X( t )=X(0)+\int_0^{ t } v(t)\mbox{d}t,
\end{equation}
and the average position becomes
\begin{equation}
\langle X( t )\rangle =X(0)+\int_0^{ t } V \mbox{d}t.
\end{equation}
Thus, the variance (\ref{variance}) is given by 
\begin{eqnarray}
F_r^2( t )&=&
\int_0^{ t }\mbox{d}t' \int_0^{ t }
\mbox{d}t'' \langle (v(t')-V)(v(t'')-V)\rangle 
\nonumber\\
&=&
2\int_0^{ t }\mbox{d}t' \int_0^{t'}\mbox{d}t'' C_v(t'-t'').
\label{frcv0}
\end{eqnarray}
Here, we assumed that the velocity correlation 
$C_v(t)=\langle (v(t_0)-V)(v(t_0+ t)-V)\rangle $ is the function of $t$
and does not depend on $t_0$.

If we take new variables $s=t'-t''$ and $s'=t'$, 
the integration (\ref{frcv0}) becomes 
\begin{eqnarray}
F_r^2( t )&=&
2\int_0^{ t }\mbox{d}s' \int_0^{s'}\mbox{d}s C_v(s)
\nonumber\\
&=&
2\left(
\left[s'\int_0^{s'}\mbox{d}s C_v(s) \right]^{s'= t }_{s'=0}
-\int_{0}^{ t }\mbox{d}s's' C_v(s')\right)
\nonumber\\
&=&
2  t \int_0^{ t }
\left(1-\frac{s}{ t }\right)C_v(s)\mbox{d}s, 
\end{eqnarray}
where we have performed the partial integration.

\subsection{Discrete time expression}

For the data at the discrete time sequence $t_j=j\tau$ ($j=0,1,2,\cdots$),
we define the velocity fluctuation as $\tilde v_j\equiv v_j-V$.
The position at time $t_n$ is given by $x_n=\sum_{i=1}^{n}v_i \tau$,
and its average is given by $\langle x_n\rangle =V n \tau$.
Thus we have
\begin{eqnarray}
\lefteqn{
\left< (x_n-\langle x_n\rangle )^2\right>
=
\left< \sum_{i=1}^{n}\tilde v_i\sum_{j=1}^{n}\tilde v_j\right>\tau^2
}\qquad
\nonumber\\
\qquad &=&
\left<
\sum_{i=1}^n {\tilde v_i}^2 +
\sum_{i=2}^n \sum_{j=1}^{i-1}\tilde v_i \tilde v_j +
\sum_{j=2}^n \sum_{i=1}^{j-1}\tilde v_i \tilde v_j 
\right> \tau^2
\nonumber\\
&=&2\sum_{i=2}^{n}\sum_{j=1}^{i-1} 
\langle \tilde v_i\tilde v_j\rangle \tau^2
+\sum_{i=1}^{n} \langle \tilde v_i^2\rangle \tau^2
\label{eq:ap2}
\end{eqnarray}
Now, we assume that $C_v(t_i,t_j)=\langle \tilde v_i \tilde v_j\rangle $
depends only on $|t_i-t_j|$, namely, $C_v(t_i,t_j)=C_v(|t_i-t_j|)$.
Then, the first term in (\ref{eq:ap2}) becomes
\begin{eqnarray}
\lefteqn{
\sum_{i=2}^{n}\sum_{j=1}^{i-1} C_v(t_i,t_j)
=
\sum_{i=2}^n \sum_{k=1}^{i-1} C_v(t_k)
}
\quad\qquad\nonumber\\
 & = &
\sum_{k=1}^{n-1} \sum_{i=k+1}^{n} C_v(t_k)
 =  
\sum_{k=1}^{n-1} (n-k) C_v(t_k)
\qquad\label{eq:ap3}
\end{eqnarray}
while we have $\sum_{i=1}^{n} \langle \tilde v_i^2\rangle =nC(0)$ for
the second term in (\ref{eq:ap2}).  Therefore, in the large $t_k$ limit,
we have from Eqs.(\ref{eq:ap2}) and (\ref{eq:ap3}) with the definition
(\ref{dmdef}) that
\begin{equation}
D = \left[
\frac{C(0)}{2}+\sum_{k=1}^{\infty}C(t_k)
\right]\tau,
\end{equation}
which is eq. (\ref{eq:discreteDm}).


\begin{thebibliography}{10}
\bibitem{com1}
The present procedure is slightly different from that by Duke in his original
  paper; Duke employed only the redistribution of population in the
  equilibration process, while we include a filament displacement in
  equilibration in accordance with the assumption that the time to achieve the
  force balance is fast in comparison with the MC step.
\end{thebibliography}

\end{document}